\documentclass[Afour,sageh,times]{sagej}

\usepackage{moreverb,url}

\usepackage{xcolor}
\definecolor{tab-blue}{HTML}{1f77b4}
\definecolor{tab-orange}{HTML}{ff7f0e}
\definecolor{tab-green}{HTML}{2ca02c}
\definecolor{tab-red}{HTML}{d62728}
\definecolor{tab-purple}{HTML}{9467bd}
\definecolor{tab-brown}{HTML}{8c564b}
\definecolor{tab-pink}{HTML}{e377c2}
\definecolor{tab-gray}{HTML}{7f7f7f}
\definecolor{tab-olive}{HTML}{bcbd22}
\definecolor{tab-cyan}{HTML}{17becf}
\definecolor{backgray}{rgb}{0.99,0.99,0.99}

\usepackage[colorlinks,bookmarksopen,bookmarksnumbered,citecolor=red,urlcolor=red]{hyperref}

\usepackage{amsmath}
\usepackage{amssymb}
\usepackage{mathtools}
\usepackage{amsthm}
\usepackage{cleveref}
\usepackage{booktabs}
\usepackage{tikz}
\usepackage{ifthen}
\usepackage{lmodern}  %
\usepackage[T1]{fontenc}  %
\usepackage{xspace}
\usepackage{listings}
\newcommand{\listingsttfamily}{\ttfamily\small}
\lstdefinestyle{prettycode}{
  basicstyle=\listingsttfamily,
  aboveskip={0.9\baselineskip},
  keepspaces=true,
  upquote=true,
  backgroundcolor=\color{backgray},
  commentstyle=\color{tab-green},
  keywordstyle=\color{tab-blue}\bfseries,
  numberstyle=\tiny\color{tab-gray},
  stringstyle=\color{tab-purple},
  breakatwhitespace=false,
  breaklines=true,
  captionpos=b,
  keepspaces=true,
  numbers=left,
  numbersep=5pt,
  showspaces=false,
  showstringspaces=false,
  showtabs=false,
  tabsize=2,
  showlines=false,
}
\theoremstyle{plain}

\theoremstyle{definition}

\theoremstyle{remark}

\newcommand{\num}[1]{{#1}}

\newcommand{\cpp}{C\texttt{++}\xspace}

\newcommand\BibTeX{{\rmfamily B\kern-.05em \textsc{i\kern-.025em b}\kern-.08em
T\kern-.1667em\lower.7ex\hbox{E}\kern-.125emX}}

\def\volumeyear{2016}
\usetikzlibrary{shapes.misc}

\begin{document}

\runninghead{Nytko et. al}

\title{Teaching An Old Dog New Tricks: Porting Legacy Code to Heterogeneous Compute Architectures With Automated Code Translation}

\author{Nicolas Nytko\affilnum{1}, Andrew Reisner\affilnum{2}, J. David Moulton\affilnum{3}, Luke N. Olson\affilnum{1}, Matthew West\affilnum{4}}

\affiliation{\affilnum{1}Department of Computer Science, University of Illinois at Urbana-Champaign.\\
\affilnum{2}Applied Computer Science, Los Alamos National Laboratory.\\
\affilnum{3}Applied Mathematics and Plasma Physics, Los Alamos National Laboratory.\\
\affilnum{4}Department of Mechanical Science and Engineering, University of Illinois at Urbana-Champaign.}

\corrauth{Nicolas Nytko, University of Illinois at Urbana Champaign.
  Department of Computer Science,
  201 North Goodwin Avenue,
  Urbana, IL,
  USA.}

\email{nnytko2@illinois.edu}

\begin{abstract}
Legacy codes are in ubiquitous use in scientific simulations; they are
well-tested and there is significant time investment in their use.  However, one
challenge is the adoption of new, sometimes incompatible computing paradigms,
such as GPU hardware.  In this paper, we explore using automated code
translation to enable execution of legacy multigrid solver code on GPUs without
significant time investment and while avoiding intrusive changes to the
codebase. We developed a thin, reusable translation layer that parses Fortran 2003 at compile time, interfacing with the existing library Loopy~\citep{KloecknerLoopy2014} to transpile to \cpp/GPU code,
which is then managed by a custom MPI runtime system that we created.
With this low-effort approach, we are able to achieve a payoff of an
approximately 2--3$\times$ speedup over a full CPU socket, and 6$\times$ in
multi-node settings.
\end{abstract}


\maketitle

\section{Introduction}

Legacy code libraries are in frequent use in scientific simulation.  In many
cases they have undergone decades of development, leading to reliable and
heavily tested portions of the overall software stack.  However, these same legacy codes pose challenges
when adapting to new hardware and when integrating with new programming
paradigms.

Here, we consider the solver package Cedar~\citep{cedar}, which is a
modern \cpp interface over the established multigrid solver code
BoxMG~\citep{DendyBoxMG1982,DendyBoxMG1983,DendyBoxMG2010,YavnehBoxMG2012,ReisnerBoxMG2018}.
BoxMG, also known as \textit{black-box} multigrid, is a Fortran code that has
been hand-optimized for execution on traditional CPU cores. To exploit further parallelism,
the codebase would need to be expanded to use GPUs and other accelerators.

For general-purpose execution on modern graphics processors, vendor-specific standards such as CUDA, HIP, and OpenCL enable
direct access to the hardware.  However, these are mostly tied to a specific architecture; if an application is written for
a single standard, then it can be challenging to expand capabilities to different or even new GPU architectures while still maintaining high performance.
For this reason, frameworks such as Kokkos~\citep{Kokkos} provide an
\emph{abstraction} over these interfaces that allows developers to follow a write-once, run-anywhere paradigm.  Even so, this still requires the nontrivial task of rewriting existing
legacy code.

To reduce the amount of work necessary to port existing legacy applications to GPUs, significant work has been invested in exploring the use of \emph{automated code translation} to accelerate
pre-existing software implementations in a nonintrusive way.  For example, Loopy~\citep{KloecknerLoopy2014,KloecknerLoopy2015Fortran} has been used to
generate performant GPU code from small Fortran-based function kernels by using loop transformation rules.  In a different way, OpenACC~\citep{OpenACC} has been proposed and used to port
large parts of existing codebases, allowing developers to tag critical sections of their codebase with compiler directives that are then used to generate GPU code by the compiler.
Works such as \citet{ChristenAutotuning2011} and \citet{HolewinskiStencil2012} have focused directly on translating stencil computations by describing them in a domain-specific language (DSL) which
is then compiled directly to machine code.  However, as pointed out by \citet{LiftStencil2018}, this requires one to implement and maintain a purpose-built optimizing compiler, which we would like to avoid;
they instead interface with a general loop-translation package~\citep{Lift2015} as we do.

In this work, we build on the automatic translation capabilities of Loopy and have implemented a tagging system similar to OpenACC that has allowed us to port our existing PDE solver
code to run on NVIDIA and AMD GPUs.  We implement our own parser and translator, as opposed to off-the-shelf solutions like OpenACC, to have more flexibility and fine-grained control over the generated code
and how it integrates with the rest of the codebase.  Additionally, we implement interprocess communication in a processor-agnostic way using the Tausch~\citep{Tausch} library, which handles
memory transfers between host and device memory when necessary (such as using GPU-aware MPI when available).

We highlight the following contributions in this work:
\begin{enumerate}
\item We develop a thin, reusable translation layer to transcribe Fortran 2003 into an intermediate GPU loop representation.
\item We apply the translation layer to the Cedar codebase, where approximately $\num{500}$ lines of code are changed, represeting only $\num{1.25}$\% of the total code.
\item We achieve approximately $\num{80} \times$ speedup as compared to the CPU code.
\item The resulting implementation is not tied to any specific GPU architecture or vendor.
\end{enumerate}

\section{Cedar and BoxMG Background}

As an application, we consider the PDE
\begin{equation}
  -\nabla \cdot ({\bf D} \nabla u) = {\bf f},
\end{equation}
and use a standard finite-difference scheme which leads to a stencil-based description of the resulting
linear system, $A\bf{x} = \bf{b}$.

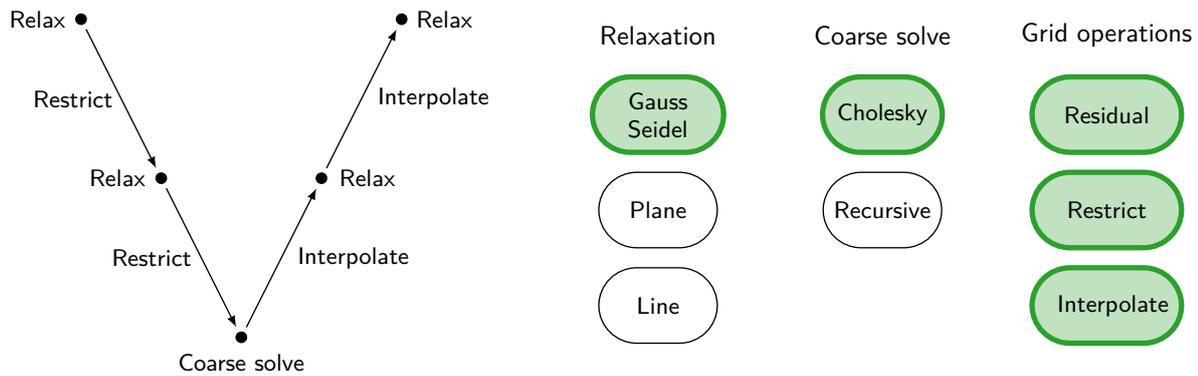
\begin{figure*}
\centering
\tikzset{every picture/.style={>=latex,line width=0.5pt,text width=1.3cm,align=center}}
\begin{tikzpicture}[x=120pt, y=120pt, font={\small \sffamily}]

  \node [circle, fill=black, scale=0.1] at (0.0, 0.0) (v1) {};
  \node [circle, fill=black, scale=0.1] at (0.25, -0.5) (v2) {};
  \node [circle, fill=black, scale=0.1] at (0.5, -1.0) (v3) {};
  \node [circle, fill=black, scale=0.1] at (0.75, -0.5) (v4) {};
  \node [circle, fill=black, scale=0.1] at (1.0, 0.0) (v5) {};

  \draw[->, shorten >= 2pt, shorten <= 2pt] (v1) to (v2);
  \draw[->, shorten >= 2pt, shorten <= 2pt] (v2) to (v3);
  \draw[->, shorten >= 2pt, shorten <= 2pt] (v3) to (v4);
  \draw[->, shorten >= 2pt, shorten <= 2pt] (v4) to (v5);

  \node[align=left, text width=4cm, font={\small \sffamily}] at (0.25, 0.0) (v1_relax) {Relax};
  \node[align=left, text width=4cm, font={\small \sffamily}] at (0.325, -0.25) (v1_restrict) {Restrict};
  \node[align=left, text width=4cm, font={\small \sffamily}] at (0.5, -0.5) (v2_relax) {Relax};
  \node[align=left, text width=4cm, font={\small \sffamily}] at (0.57, -0.75) (v2_restrict) {Restrict};
  \node[align=center, text width=4cm, font={\small \sffamily}] at (0.5, -1.08) (v3_solve) {Coarse solve};
  \node[align=left, text width=4cm, font={\small \sffamily}] at (1.15, -0.75) (v3_interp) {Interpolate};
  \node[align=left, text width=4cm, font={\small \sffamily}] at (1.28, -0.5) (v4_relax) {Relax};
  \node[align=left, text width=4cm, font={\small \sffamily}] at (1.4, -0.25) (v3_interp) {Interpolate};
  \node[align=left, text width=4cm, font={\small \sffamily}] at (1.52, 0.0) (v5_relax) {Relax};

  \def\mh{1cm}

  \def\sx{1.8}
  \def\sy{-0.3}
  \node [rounded rectangle, draw=tab-green, line width=2pt, fill=tab-green, fill opacity=0.3, text opacity=1, minimum height=\mh] at (\sx, \sy) (gauss) {Gauss Seidel};
  \node [rounded rectangle, draw, minimum height=\mh] at (\sx, {\sy - 0.3}) (plane_relax) {Plane};
  \node [rounded rectangle, draw, minimum height=\mh] at (\sx, {\sy - 0.6}) (kacz) {Line};
  \node[align=center, text width=4cm, font={\sffamily}] at ({\sx}, {\sy + 0.25}) (relax_kernels) {Relaxation};

  \def\sx{3.2}
  \def\sy{-0.3}
  \node [rounded rectangle, draw=tab-green, line width=2pt, fill=tab-green, fill opacity=0.3, text opacity=1, minimum height=\mh, minimum width=2.2cm] at (\sx, \sy) (residual) {Residual};
  \node [rounded rectangle, draw=tab-green, line width=2pt, fill=tab-green, fill opacity=0.3, text opacity=1, minimum height=\mh, minimum width=2.2cm] at (\sx, {\sy - 0.3}) (restriction) {Restrict};
  \node [rounded rectangle, draw=tab-green, line width=2pt, fill=tab-green, fill opacity=0.3, text opacity=1, minimum height=\mh, minimum width=2.2cm] at (\sx, {\sy - 0.6}) (interpolation) {Interpolate};
  \node[align=center, text width=4cm, font={\sffamily}] at ({\sx}, {\sy + 0.25}) (gridop_kernels) {Grid operations};

  \def\sx{2.5}
  \def\sy{-0.3}
  \node [rounded rectangle, draw=tab-green, line width=2pt, fill=tab-green, fill opacity=0.3, text opacity=1, minimum height=\mh] at (\sx, \sy) (cholesky) {Cholesky};
  \node [rounded rectangle, draw, minimum height=\mh] at (\sx, {\sy - 0.3}) (lu) {Recursive};
  \node[align=center, text width=4cm, font={\sffamily}] at ({\sx}, {\sy + 0.25}) (cs_kernels) {Coarse solve};
\end{tikzpicture}
\caption{A flow chart of the various kernels that are executed for an example V-cycle multigrid solver.  While there are many combinations of kernels to be run, only a subset are needed for any particular problem.  An example of a specific run configuration is highlighted with bold outline.}\label{fig:vcycle_flowchart}
\end{figure*}

To solve the linear system, we use a stencil-based multigrid method based on
Blackbox multigrid (BoxMG)~\citep{DendyBoxMG1982,DendyBoxMG1983,DendyBoxMG2010}.
BoxMG is a multigrid method that implements a sequence of relaxation
(Gauss-Seidel) and restriction/interpolation operatorations to iteratively
correct the error in a solution on coarser
levels~---~see~\cref{fig:vcycle_flowchart}.  When traversing the grid hierarchy
to the coarsest level, a direct solver is used (on a small system). BoxMG
implements relaxation, restriction, interpolation, and the construction of
coarse programs through local stencil operations; details can be found in
~\citet{DendyBoxMG2010} and \citet{ReisnerBoxMG2018}.

In addition to BoxMG, the software layer Cedar provides a high level
abstraction that handles memory management at scale, \cpp support, MPI
communication, and performance optimizations for operations such as plane
relaxation~\citep{ReisnerBoxMG2018,reisnerplane}.

\begin{figure}
\centering
\begin{lstlisting}[language=Fortran]
!#LOOPY_START(assume="nxc>=3")
DO jc=2, Nyc-1
  DO ic=2, Nxc-1
    j = jstart + (jc-1)*2
    i = istart + (ic-1)*2
    QC(ic,jc) = &
      Ci(ic,jc,LNE) * Q(i-1,j-1)
    + Ci(ic,jc,LA) * Q(i,j-1) &
    + Ci(ic+1,jc,LNW) * Q(i+1,j-1) &
    + Ci(ic,jc,LR) * Q(i-1,j) &
    + Q(i,j) &
    + Ci(ic+1,jc,LL) * Q(i+1,j) &
    + Ci(ic,jc+1,LSE) * Q(i-1,j+1) &
    + Ci(ic,jc+1,LB) * Q(i,j+1) &
    + Ci(ic+1,jc+1,LSW) * Q(i+1,j+1)
  ENDDO
ENDDO
!#LOOPY_END
\end{lstlisting}
\caption{A Fortran code snippet showing the kernel that computes restriction.  Here, \texttt{QC} contains the output coarse grid values, \texttt{Q} is the input fine-grid, and \texttt{Ci} is the interpolation stencil.
  The codebase has well-structured, stateless loops.}\label{fig:restrict_kernel}
\end{figure}

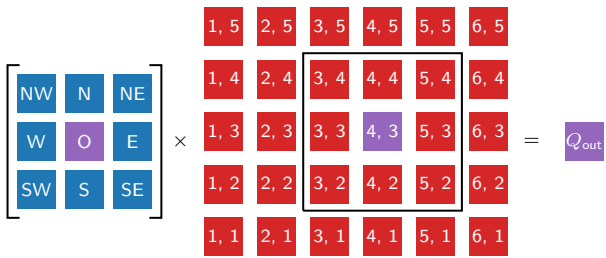
\begin{figure}
\centering
\definecolor{tab-blue}{HTML}{1f77b4}
\definecolor{tab-orange}{HTML}{ff7f0e}
\definecolor{tab-green}{HTML}{2ca02c}
\definecolor{tab-red}{HTML}{d62728}
\definecolor{tab-purple}{HTML}{9467bd}
\definecolor{tab-brown}{HTML}{8c564b}
\definecolor{tab-pink}{HTML}{e377c2}
\definecolor{tab-gray}{HTML}{7f7f7f}
\definecolor{tab-olive}{HTML}{bcbd22}
\definecolor{tab-cyan}{HTML}{17becf}
\begin{tikzpicture}[x=18pt, y=18pt, font={\scriptsize \sffamily}]
  \def\w{3.2}
  \def\h{0.25}
  \def\sx{0.4}
  \def\sy{0.4}
  \draw[line width=0.75pt] (\sx+\h, \sy) -- (\sx, \sy) -- (\sx, \sy+\w) -- (\sx+\h, \sy+\w); %
  \draw[line width=0.75pt] (\sx+\w-\h, \sy) -- (\sx+\w, \sy) -- (\sx+\w, \sy+\w) -- (\sx+\w-\h, \sy+\w); %

  \def\h{0.8}
  \fill[fill=tab-blue] (0.6, 0.6) rectangle ++ (\h, \h) ++(-\h/2, -\h/2) node[text=white] {SW};
  \fill[fill=tab-blue] (1.6, 0.6) rectangle ++ (\h, \h) ++(-\h/2, -\h/2) node[text=white] {S};
  \fill[fill=tab-blue] (2.6, 0.6) rectangle ++ (\h, \h) ++(-\h/2, -\h/2) node[text=white] {SE};
  \fill[fill=tab-blue] (0.6, 1.6) rectangle ++ (\h, \h) ++(-\h/2, -\h/2) node[text=white] {W};
  \fill[fill=tab-purple] (1.6, 1.6) rectangle ++ (\h, \h) ++(-\h/2, -\h/2) node[text=white] {O};
  \fill[fill=tab-blue] (2.6, 1.6) rectangle ++ (\h, \h) ++(-\h/2, -\h/2) node[text=white] {E};
  \fill[fill=tab-blue] (0.6, 2.6) rectangle ++ (\h, \h) ++(-\h/2, -\h/2) node[text=white] {NW};
  \fill[fill=tab-blue] (1.6, 2.6) rectangle ++ (\h, \h) ++(-\h/2, -\h/2) node[text=white] {N};
  \fill[fill=tab-blue] (2.6, 2.6) rectangle ++ (\h, \h) ++(-\h/2, -\h/2) node[text=white] {NE};

  \def\is{4}
  \def\js{3}

  \def\sx{3.4}
  \def\sy{-1.5}
  \def\h{0.8}
  \def\hs{1.1}
  \foreach \i in {1,...,6} {
    \foreach \j in {1,...,5} {
      \ifthenelse{\i=\is \AND \j=\js}{
        \fill[fill=tab-purple] (\sx+\i*\hs, \sy+\j*\hs) rectangle ++(\h,\h) ++(-\h/2, -\h/2) node[text=white] {\i, \j};
      }{
        \fill[fill=tab-red] (\sx+\i*\hs, \sy+\j*\hs) rectangle ++(\h,\h) ++(-\h/2, -\h/2) node[text=white] {\i, \j};
      }
    }
  }

  \def\hd{(\hs-\h)/2}
  \draw[line width=0.75pt]
  ({\sx + (\is-1)*\hs - \hd}, {\sy + (\js-1)*\hs - \hd}) --
  ({\sx + (\is+2)*\hs - \hd}, {\sy + (\js-1)*\hs - \hd}) --
  ({\sx + (\is+2)*\hs - \hd}, {\sy + (\js+2)*\hs - \hd}) --
  ({\sx + (\is-1)*\hs - \hd}, {\sy + (\js+2)*\hs - \hd}) --
  ({\sx + (\is-1)*\hs - \hd}, {\sy + (\js-1)*\hs - \hd});

  \def\w{3.2}
  \def\h{0.25}
  \def\sx{0.4}
  \def\sy{0.4}

  \def\h{0.8}
  \fill[fill=tab-purple] (12.0, 1.6) rectangle ++ (\h, \h) ++(-\h/2, -\h/2) node[text=white] {$Q_{\text{out}}$};

  \node at (4, 2) {$\times$};
  \node at (11.3, 2) {$=$};

\end{tikzpicture}
\caption{BoxMG kernels are stencil-based and operate on a structured 2D array, similar to convolution in image processing.  However, the stencil may be different at each point and so extra data transfers are incurred as compared to regular convolution.  In this example, a weighted affine sum of the neighbors of point \texttt{4,3} are computed and stored as $Q_{\text{out}}$.}\label{fig:stencil_operator}
\end{figure}

\subsection{Code Structure}

Cedar provides high-level abstraction to the BoxMG code by wrapping
primitive Fortran datatypes and subroutines with \cpp to provide type safety,
memory management, communication, and a generally simpler-to-use interface. The
underlying solver code is mostly intact, and is compiled directly from the
Fortran code and linked to the rest of the Cedar interface.  The goal of this
work is to seemlessly extend the execution of Cedar to GPUs, without extensive
writing (or rewriting) of code.

We use the term \emph{kernel} in the context of the BoxMG codebase to refer to
a specific implementation of a key functionality of the solver; for example,
performing interpolation or relaxation on a specific coarse grid level are
both kernels.  The existing solver kernels have a structure
that is well suited for automatic translation to GPU code, namely:
\begin{enumerate}
\item no memory allocations are performed within kernels,
\item no global state is used within kernels,
\item kernel loops have well-defined bounds that do not change between iterations, and
\item kernels contain few function calls to other kernels, and no recursion in general. \label{item:function_calls}
\end{enumerate}

For the kernel-to-kernel calls exposed in this work, these only represent calls
into MPI for the distributed memory parallel portions of the codebase or calls to BLAS/LAPACK for the coarsest-level solve.  We
translate the function calls with a small amount of wrapper code; more
details about our approach are provided in the MPI subsection of our code
translation implementation.

\section{Code Translation}

\begin{figure}
\centering
\tikzset{every picture/.style={>=latex,line width=0.5pt,text width=1.3cm,align=center}}

\begin{tikzpicture}[x=100pt, y=100pt, font={\scriptsize \sffamily}]

\fill[fill=tab-orange, fill opacity=0.15] (-1, -1.5) rectangle (0.5, 0.3);

\node [circle, draw, fill=white] at (0.0,  0.0) (parse) {Parse Fortran};
\node [circle, draw, fill=white] at (0.0,  -0.6) (cpp) {C++ IR};
\node [circle, draw, fill=white] at (-0.7,  -0.6) (loopy) {Loopy IR};
\node [circle, draw, fill=white] at (0, -1.2) (host) {Host Compilation};

\draw[line width=0.75pt] (0.55, 0.25) -- (0.6, 0.25) -- (0.6, -1.45) -- (0.55, -1.45);
\node[align=left, text width=4cm, font={\small \sffamily}] at (1.2, -0.6) (compile_time) {Compile time};

\fill[fill=tab-blue, fill opacity=0.15] (-1, -1.5) rectangle (0.5, -2.7);
\node [circle, draw, fill=white] at (0, -1.8) (device_gen) {Device Code Gen.};
\node [circle, draw, fill=white] at (-0.7, -1.8) (device_comp) {Device Compilation};

\node [circle, draw, fill=white] at (0, -2.4) (exec) {Host/Device Execution};

\draw[line width=0.75pt] (0.55, -1.55) -- (0.6, -1.55) -- (0.6, -2.65) -- (0.55, -2.65);
\node[align=left, text width=4cm, font={\small \sffamily}] at (1.2, -2.1) (run_time) {Run time};

\draw[->] (parse) to (cpp);
\draw[->] (parse) to (loopy);
\draw[->] (cpp) to (host);

\draw[->] (loopy) to (device_gen);
\draw[->] (device_gen) to (device_comp);
\draw[->] (device_comp) to (exec);
\draw[->, rotate=90] (-1.2, -0.24) parabola bend(-1.8, -0.4) (-2.4, -0.24);

\end{tikzpicture}
\caption{The translation steps and execution sequence.  Host-side code is translated at compile time, while device-side code is translated into an intermediate representation and compiled at run-time.}\label{fig:runtime_workflow}
\end{figure}
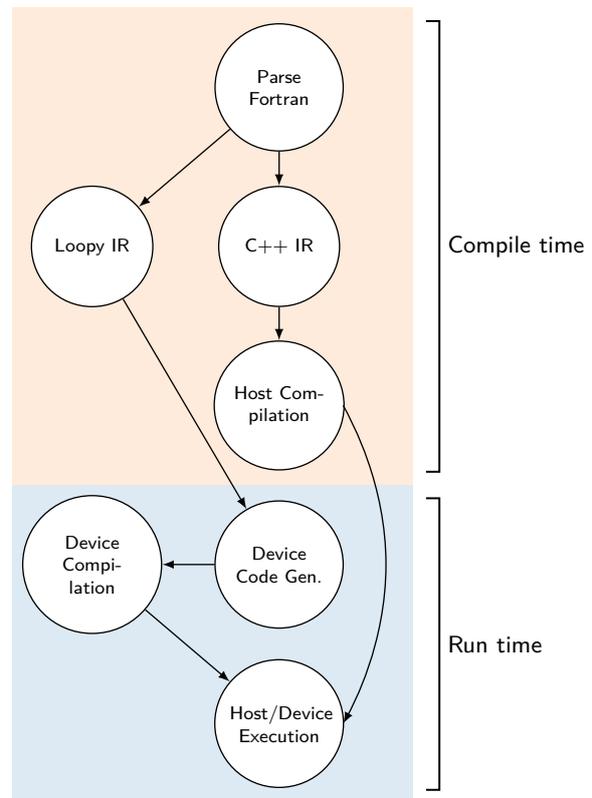

While positioning the codebase for execution on GPUs, one of our main goals is to
minimize the amount of existing code that is removed or changed.  To this end,
we make use of \textit{code translation} to automatically generate GPU code.
Our approach to code translation consists of two parts:
\begin{enumerate}
\item a compile-time Fortran parser, and
\item a run-time system that interfaces with the GPU and handles data transfers.
\end{enumerate}
These are also enumerated in \cref{fig:runtime_workflow}.

One advantage of this approach to GPU execution is a separation of concerns:
keeping the legacy codebase undisturbed, while using additional tools to parse
and translate (and test).  As a result, the performance aspects of the GPU can
be focused on the transformation tool.  There are several
code translation tools available.  In this work, we use
Loopy~\citep{KloecknerLoopy2014} to generate high-performance GPU code.  At a
high level, Loopy defines a model for defining code that operates on arrays.
The bounds for the arrays, as well as the operations to be performed, are specified and
the framework generate device code to manipulate arrays in memory.
In addition it provides transformations such as loop tiling, which we use to
gain parallelism on the GPU.

\subsection{Parsing}

The parsing of the existing Fortran code is performed at compilation time of the Cedar codebase using a custom CMake rule.  While the Loopy framework has support
for Fortran parsing~\citep{KloecknerLoopy2015Fortran}, it is intended for fully translating small routines, instead of the large pieces of code
found in BoxMG\@.  In addition, conditional statements are completely unsupported, which motivated our use of a custom parser.

To mark portions of the existing code to be parallelized for execution on the GPU, we employed a \emph{tagging} system which can be seen in~\cref{fig:restrict_kernel}.  The section of code is enclosed with the lines \texttt{!\#LOOPY\_START}
and \texttt{!\#LOOPY\_END} to signify the start and end of the parallel section.  As these are valid Fortran comments, the same code compiles even without use of the parser (i.e., on a standard Fortran compiler).  Our parser then reads the Fortran code, and performs
one of two behaviors depending on whether a section of code is marked as a parallel section.  First, if a section of code \emph{is not} marked parallel, this it is assumed to be \emph{host-specific} code, and the code is not transpiled to \cpp.
To allow for improved interoperability with Cedar (such as typing and opaque data transfers; more below), we transpile to \cpp rather than directly compiling to machine code.
Second, if a section of code \emph{is} marked to be parallel, then an intermediate representation of Loopy instructions is generated,
including loop domains and loop behavior.  Our logic for determining loop domains is simple: if multiple nested loops are encountered inside a parallel section, then each inner loop iteration is executed on a single GPU thread.  This is sufficient for our use case; however, if more complex behaviors are needed they can be added as flags in the tagged section.

\begin{figure}
  \centering
  \begin{minipage}{0.45\linewidth}
  \begin{lstlisting}[language=Fortran]
I = 2
J = 2
DO IC=3, NC
  I = I + 2
  DO JC=3, NC
    J = J + 2
    QF(I,J) = ...
  ENDDO
ENDDO\end{lstlisting}
  \end{minipage}
  \hfill
  \begin{minipage}{0.45\linewidth}
  \begin{lstlisting}[language=Fortran]


DO IC=3, NC
  DO JC=3, NC
    I = (IC-1) * 2
    J = (JC-1) * 2
    QF(I,J) = ...
  ENDDO
ENDDO\end{lstlisting}
  \end{minipage}
  \caption{Example code snippet with dependencies present between loop iterations (left), and the same snippet rewritten to remove such dependencies (right).  In both listings, \texttt{QF} is an array being accessed by the indices \texttt{I} and \texttt{J}.}
  \label{fig:loop_indices}
\end{figure}

With this approach, only small amounts of code are modified in order for the system to automatically generate parallel GPU code.  For example, in the existing
serial codebase many loops are written with dependencies present between loop iterations, mostly when array indices are computed (see \cref{fig:loop_indices}).
As such, the loop dependencies are rewritten and subsequently tested to ensure the behavior is not changed for both the CPU and GPU implementations.

\subsection{Runtime}
We have written a reusable runtime that interfaces with the GPU and handles operations such as transferring data when appropriate and interfacing with MPI\@.  On application startup, a GPU vendor interface selects OpenCL or CUDA based on the detected hardware,
and the device initializes the driver.  Then, all Loopy kernels are loaded and compiled for the specific GPU implementation.  Kernels are compiled at run-time, allowing the selection of different interfaces.  Additionally,
the compiled output is cached, so a performance hit is only incurred on the first launch of the application.

To add an abstraction to data ownership, the runtime provides an abstract \emph{Buffer type}, which owns a specific piece of data on either the host, the device, or both.  This buffer type tracks the reads and writes, and performs data transfers automatically.
For example, a write by the host CPU followed by a read from the device GPU will incur a data transfer from CPU to GPU, however multiple successive reads from the GPU will not transfer data (apart from potentially the first read, if the data is not already present on the GPU).  This is implemented
internally using lazy evaluation of buffer accesses with a reference \cpp accessor class: when the buffer is indexed on the CPU a reference to the piece of memory is returned.  If the reference is read from or written to, then dirty bits on the underlying buffer are checked and data transfers
are performed.  One benefit of the lazy evaluation is that it allows simple translation of Fortran-style passing of references into existing arrays.  Additionally, the read/write tracking can be ``batched'' if, for example, the Fortran parser detects that multiple reads or writes
will be performed by the same device, to reduce performance penalties.

\subsection{MPI Calls}

Function calls with the Fortran \texttt{CALL} keyword are translated directly to equivalent \cpp function calls.  The translation is trivial when the calling convention
is identical~---~e.g., BoxMG calls other BoxMG routines.  However, additional steps are needed to translate code
that contains MPI function calls due to the different calling conventions present between C and Fortran.  In these situations, wrapper implementations around the MPI
calls are defined to correct the misalignment in the two calling conventions.  Data-types are converted between the two language specifications using
conversion routines defined by MPI, and the respective C implementation is called.  An example is given in \cref{fig:mpi_recv}.
\begin{figure}
  \begin{lstlisting}[language=C]
void MPI_Recv(
  void* buf, int count,
  MPI_Datatype datatype,
  int source, int tag,
  MPI_Fint comm_f, MPI_Status& status,
  MPI_Fint& ierr) {

  MPI_Comm comm_c=MPI_Comm_f2c(comm_f);
  ierr=MPI_Recv(buf, count, datatype,
    source, tag, comm_c, &status);
}\end{lstlisting}
  \caption{Example wrapper implementation for the \texttt{MPI\_Recv} function.  The Fortran handle is converted to a C communicator type,
    then the respective C routine is called.}
  \label{fig:mpi_recv}
\end{figure}

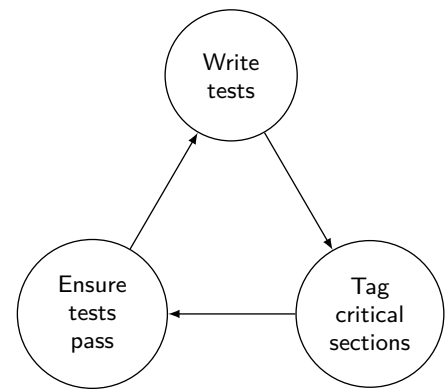
\begin{figure}
\centering
\tikzset{every picture/.style={>=latex,line width=0.5pt,text width=1.3cm,align=center}}

\begin{tikzpicture}[x=120pt, y=120pt, font={\small \sffamily}]
\node [circle, draw] at ( 0.0  ,   0.5) (tests) {Write tests};
\node [circle, draw] at ( 0.433,  -0.25) (tag) {Tag critical sections};
\node [circle, draw] at (-0.433,  -0.25) (pass) {Ensure tests pass};

\draw[->] (tests) to (tag);
\draw[->] (tag) to (pass);
\draw[->] (pass) to (tests);
\end{tikzpicture}
\caption{The steps required to translate serial host code into device code.  Tests should be written to ensure correctness of both existing and translated code.}\label{fig:translation_workflow}
\end{figure}

\section{Timing Results}

In this section, we show timing results for a direct use case~---~i.e., small amounts of optimization were performed to minimize data transfers, but extensive tuning
to maximize the FLOP rate of the solver on the GPU was avoided.  In this example, cuBLAS~\citep{cuBLAS} is used for several basic linear algebraic routines, for example in the $\ell^2$ norm and to perform a Cholesky factorization/solve on the coarse
level; all other parts of the solver stack are directly translated from the original Fortran implementation.  Performance tests are on the GPU nodes on NCSA Delta:
each compute node consists of one 64-core AMD Milan processor and 4 Nvidia A100 GPUs.  These nodes have 256~GiB of total system ram available to the CPU and 40~GiB of high-bandwidth memory per GPU.  Nodes are connected with a 200~Gb/s HPE Slingshot interconnect.

\begin{figure}
\centering
\includegraphics{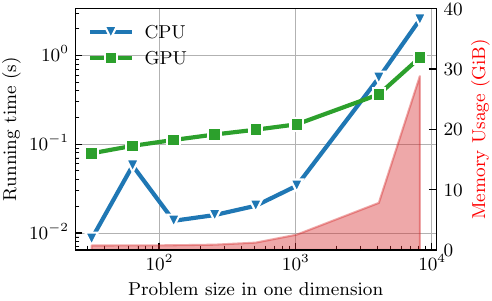}
\caption{Run times for the solver on a single CPU socket (64 cores) and a single GPU.}\label{fig:timing_serial}
\end{figure}

We measure performance of the solver running on a single GPU versus a full CPU socket (64 cores).  The problem size is increased until the problem and auxiliary solver data fully saturate the GPU memory (40~GiB), as seen in~\cref{fig:timing_serial}.  Here we observe that there is little benefit to using the GPU over the CPU until the problem size reaches $\num{4096}$ in one dimension. At this size, we see a $1.5\times$ and $2.6\times$
speedup on the GPU for the $\num{4096}$ and $\num{8192}$ sized problems.  Indeed, smaller problems are more sensitive to the overhead of launching compute kernels on the GPU, which becomes more costly than the actual
time spent in computation on the GPU.

\begin{figure}
\centering
\includegraphics{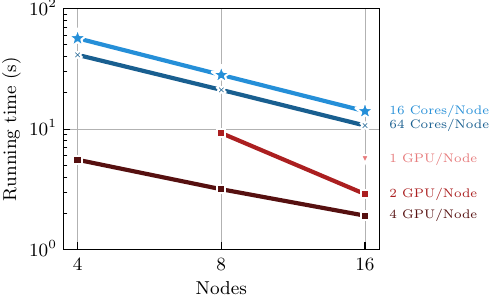}
\caption{Run times for the solver when varying the number of cores, keeping the total degrees of freedom fixed to $N=\num{32768}^2$.}\label{fig:timing_mpi}
\end{figure}

Next, we perform a weak scaling test with MPI, using 4, 8, and 16 compute nodes to solve a problem of global size $\num{32768}$ in each dimension.  This scaling test is executed on both the CPU and GPU implementations of the solver:
in separate tests, the CPU sockets are fully saturated and all 4 GPUs are fully saturated to run the solver.  The results of these tests are summarized in \cref{fig:timing_mpi}.  We see that memory becomes a limiting factor:
in the case of 1 or 2 GPUs per node, the subproblems are unable to fit into memory unless many nodes are used.  However, whenever the GPU is used a speedup is obtained over any of the CPU tests, yielding a
notable speedup when 4 GPUs/node are used~---~a 6$\times$ speedup is obtained when 16 nodes are used.

\begin{figure*}
\centering
\includegraphics{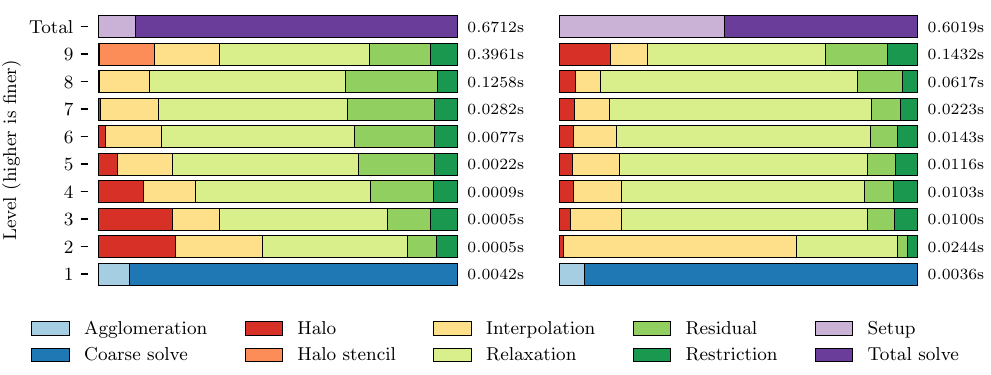}
\caption{Per-kernel time for $N=\num{1024}^2$, on the CPU (left) and GPU (right).  Overall setup and solve time are displayed on the top (``Total'') bar.  Collective time spent on each individual level is displayed to the right of each bar.}\label{fig:kernel_timings}
\end{figure*}

Finally, we time individual kernels on each level of the solver hierarchy (see \cref{fig:kernel_timings}).
For levels where the problem is sufficiently large (7--9), we see that the GPU has enough parallelism to exploit that we gain a speedup when compared to the CPU.  However,
for coarser levels (3--6), the overall run time is dominated by the overhead of launching GPU kernels; this is further corroborated by noting that each level on the
GPU takes roughly the same amount of computation time, while we see the expected decrease on the CPU\@.

\section{Conclusions}

We used automatic code translation to enable GPU acceleration of legacy, stencil-based Fortran code.  While there was a development cost in writing the parser and translation runtime,
we were able to accelerate an existing codebase with minimal intrusive changes.  Morever, the parser and translation code is is reusable and can be applied to other projects.

By accelerating our existing PDE solver code, we were able to achieve an approximately 2--6$\times$ speedup compared to running the solver on the CPU, without any special performance optimizations and without substantial changes
to the existing codebase.

There are multiple further avenues of work to explore.  A hybrid CPU-GPU solve, where the GPU is used for the fine levels of computation and the CPU is used
for coarser levels once there is insufficient parallelism in the problem size to exploit, may be beneficial.  There is also the possibility of further tuning the generated stencil computations, perhaps by using general techniques like temporal blocking~\citep{TemporalBlocking2015}.  The possibility of introducing more data reuse
could also potentially give speedups in the solver; our performance is mainly bounded by memory bandwidth of the GPU.  By generating multiple (block) initial solution vectors and solving in a block fashion,
similarly to what is done in \citet{Moufawad2020}, an accurate solution could potentially be extracted from the block solutions in fewer overall iterations.

\begin{acks}
  This work used the Delta system at the National Center for Supercomputing Applications through allocation CIS230390 from the Advanced Cyberinfrastructure Coordination Ecosystem: Services \& Support (ACCESS) program, which is supported by National Science Foundation grants \#2138259, \#2138286, \#2138307, \#2137603, and \#2138296.

  This material is based in part upon work supported by the Department of Energy, National Nuclear Security Administration, under Award Number DE-NA0003963.
\end{acks}

\bibliographystyle{SageH}
\bibliography{paper_boxmg_gpu_strip}

\end{document}